\documentclass[aps,prb,twocolumn,showpacs,groupedaddress]{revtex4}
\usepackage{graphicx}
\usepackage{amsmath,amssymb}

\begin{document}
\bibliographystyle{apsrev}

\title{Nb/CuNi Sandwiches as Superconducting Spin-Valve Core Structures}
\author{A. Potenza}
\affiliation{School of Physics and Astronomy, E. C. Stoner
Laboratory, University of Leeds, Leeds. LS2 9JT United Kingdom}
\author{C. H. Marrows}
\email[Electronic mail: ]{c.marrows@leeds.ac.uk}
\affiliation{School of Physics and Astronomy, E. C. Stoner
Laboratory, University of Leeds, Leeds. LS2 9JT United Kingdom}

\date{\today}

\begin{abstract}
We have investigated CuNi/Nb/CuNi trilayers, as have been recently
used as the core structure of a spin-valve like device [J. Y. Gu et
al., Phys. Rev. Lett. {\bf 89}, 267001 (2002)] to study the effect
of magnetic configurations of the CuNi layers on the critical
temperature, $T_{\rm C}$, of the superconducting Nb. After
reproducing a $T_{\rm C}$ shift of a few mK, we have gone on to
explore the performance limits of the structure. The results showed
the $T_{\rm C}$ shift we found to be quite close to the basic limits
of this particular materials system. The ratio between the thickness
and the coherence length of the superconductor and the interfacial
transparency were the main features limiting the $T_{\rm C}$ shift.
\end{abstract}
\maketitle


Superconductor (S)/ferromagnet (F) proximity systems show many
interesting physical effects originating from the coexistence of
these two mutually exclusive orderings of matter \cite{Chien}.
Ferromagnetism is expected to suppress superconductivity, as the
presence of an exchange field breaks the time reversal symmetry of a
Cooper pair. However, the splitting of the energy levels of the
spin-$\uparrow$ and spin-$\downarrow$ electrons in the pair may not
totally suppress the superconducting state. Exotic superconducting
states, e.g. the Larkin-Ovchinnikov-Fulde-Ferrell (LOFF) state
\cite{Fulde,Larkin}, can be adopted. In S/F multilayers, the mutual
coexistence of the two effects is responsible for re-entrant
superconductivity in the critical temperature ($T_{\rm C}$) versus F
thickness ($d_{\rm m}$) behavior \cite{Strunk,Jiang}, and
$\pi$-junction effects \cite{Kontos,Rusanov}. Particularly
interesting phenomena are predicted to happen in the limit of thin S
layers, when the thickness $d_{\rm s}$ is comparable to the
coherence length $\xi_{\rm s}$. Non-local effects due to spatial
variations in the exchange field over the coherence length will
strongly affect the superconducting state. Therefore, it will be
possible for a SC to distinguish between parallel (P) or
antiparallel (AP) magnetic configurations in two adjacent F layers
\cite{buzdinepl,melinepl}. In the latter case, the opposite
orientations should partially cancel out, resulting in a much weaker
effective field acting on the pair, and consequently a higher
$T_{\rm C}$.

This idea was the basis of a proposed model for a so-called
superconducting spin-valve \cite{Tag1}, an F/S/F layer sequence
where it is possible to switch between P and AP alignment of the two
F layer moments. If the shift in critical temperature,
$\Delta{T_{\rm C}}$, between the P ($T_{\rm C,P}$) and AP ($T_{\rm
C,AP}$) configurations is bigger than the superconducting transition
width, $\Delta{W}$, then for $T_{\rm C,P}<T<T_{\rm C,AP}$, it is
possible to valve the supercurrent by applying only a relatively
small magnetic field to switch a soft F layer.

A first realization such a spin-switch has been reported, using CuNi
for the F layers and Nb for the S spacer \cite{Gu}. Thin S layers
are required, so the use of Nb and CuNi seems promising because of
the very weak ferromagnetism of CuNi, which can allow the S layer to
be far thinner if compared to systems with stronger ferromagnets
such as Fe or Co. The device showed a small but measurable effect,
with $\Delta T_{\rm C} \approx 6$ mK, but this is still much less
than $\Delta W \approx 0.1$ K. Very recently this group has
published a theoretical description of this structure
\cite{you2004}. In the present Communication, we begin by reporting
our results from a nominally identical sample, which reproduces just
what was found in Ref. \onlinecite{Gu}, confirming the existence of
the superconducting spin-valve effect. Further results are then
reported on the properties of Nb/CuNi trilayers, and the intrinsic
limits in the device performance in this materials system are found
by constraining as many parameters as possible experimentally,
entirely using material grown in the same sputter chamber.

The samples, with planar dimensions of 10 mm $\times$ 2 mm, were
grown by dc magnetron sputtering on Si(100) substrates. The CuNi was
sputtered from an alloy target, and characterized by vibrating
sample magnetometry (VSM). The Curie temperature of a 5000 \AA\
thick film, determined as the temperature where there is an upturn
in the value of magnetic moment, was found to be $\sim40$ K, which
corresponds to a composition of $\sim 51$ atomic per cent Ni
\cite{Stampe}, which compares well with $53 \pm 2$ atomic per cent
Ni, determined by x-ray photoelectron spectroscopy. In the first set
of samples, S1, spin-valve structures were grown with the layer
sequence
Ta/Ni$_{80}$Fe$_{20}$/CuNi/Nb/CuNi/Ni$_{80}$Fe$_{20}$/FeMn/Ta and
the Nb thickness $d_{\rm s}$ varied, and 100 \AA\ CuNi layers used.
In S1, the core trilayer structure was embedded within two 30 \AA\
Ni$_{80}$Fe$_{20}$ layers, as in Ref. \onlinecite{Gu}, in order to
improve the switching characteristics of the CuNi. The second set,
S2, is a sequence of CuNi layers with thickness $d_{\rm m}$
increasing from 25 to 800 \AA. The third set, S3, is composed of
CuNi(100 \AA)/Nb/CuNi(100 \AA) trilayer structures with varying Nb
layer thickness, and a Ta cap of $\sim$30 \AA. The last set, S4,
consists of similar trilayers to S3, but with a Nb thickness of 230
\AA\, varying CuNi thickness, and $\sim$50 \AA\ of Ge as a capping
layer.

\begin{figure}
  \includegraphics[width=8.0cm]{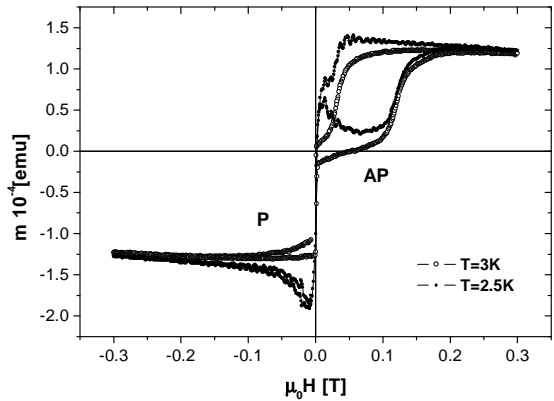}
\caption{Hysteresis loops for the 180\AA\ Nb sample of the S1 set
at temperatures just above (3 K) and just below (2.5 K) the
critical temperature ($T_{\rm C} \approx 2.8$ K).} \label{F1}
\end{figure}

In Fig. \ref{F1} the hysteresis curves for the sample of S1 with
$d_{\rm s}$=180 \AA\ are displayed for $T=3$ K and $T=2.5$ K, either
side of the superconducting transition. At 3 K the magnetization
reproduces a typical spin-valve behavior, where a region of AP
alignment of the magnetic moments in the two F layers is
distinguishable for $0<\mu_0 H< 0.25$ T on the forward going branch
of the curve.

\begin{figure}
  \includegraphics[width=8.0cm]{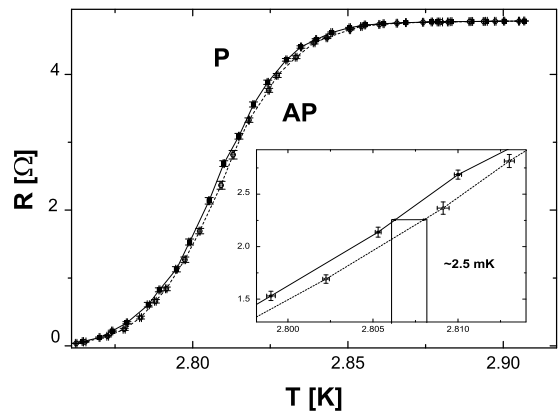}
\caption{Resistance vs. $T$ plot close to $T_{\rm C}$ for the same
sample (S1) as Fig. \ref{F1}. The upper curve refers to P
alignment ($-30$ mT), while the lower is for the AP case ($+30$
mT). The inset is a magnification showing a $T_{\rm C}$ shift of
$\sim 2.5$ mK.} \label{F2}
\end{figure}

The electrical resistance of the sample was measured by means of a
low temperature dc 4-point probe in the $\emph{in-plane}$ geometry
(probe spacing of 2.5 mm), where the external field is collinear to
the sample length. Prior to the measurement for the superconducting
transition (Fig. \ref{F2}), a magnetic field of $+0.5$ T was
applied. Subsequent application of small positive/negative external
magnetic fields corresponds to switching between AP/P configurations
by reversing the free layer only, in the usual spin-valve scheme.
The resistance was then measured at alternating fields of $\pm 30$
mT as the sample was cooled through the superconducting transition.
Since these fields were equal and opposite, the direct effect of the
applied field on the S layer is canceled out, as well as other
effects of even symmetry such as magnetostriction in the F layers.
We can see from Fig. \ref{F1} that our measurement fields are far
from the coercivities of both magnetic layers, so that we can be
confident that domain effects such as exchange field averaging
\cite{RusanovPRL2004} and spontaneous vortex formation
\cite{Ryzanov2003} do not play any significant role. An averaging
process over many consecutive readings was carried out, because of
the changes of the resistance due to thermal fluctuations across the
transition, reflected in the displayed error bars. In Fig. \ref{F2}
the resistance vs. $T$ plot for the sample shows the $T_{C}$ split
obtained by switching between P and AP alignment on F layers. The
thickness of the Nb layer is nominally identical to the sample of
Ref. \onlinecite{Gu} and the data are indeed very similar, showing a
$T_{\rm C} = $2.82 K, a transition width $\Delta{W}\approx 0.1$ K
and a $\Delta T_{\rm C} \approx 2.5$ mK.

Theoretical predictions of $\Delta T_{\rm C}$ are usually orders of
magnitude larger. In order to find out more quantitatively the main
limitations on performance, we carried out a series of experiments
that were interpreted in terms of a theoretical model for the
proximity effect by Tagirov \cite{Tag1}. Although this theory treats
singlet pairing only, this is all that is required to deal with the
collinear P and AP configurations that we concern ourselves with
here \cite{fominov2003}. This theory allows the calculation of the
$T_{\rm C}$ of the F/SC/F trilayer (using Eqns. 8 and 9 of Ref.
\onlinecite{Tag1} for the P state, and 8 and 10 for the AP state)
based on the values of four parameters:
\begin{equation}
  \varepsilon=\sqrt{6}\frac{N_{\rm s}\xi_{\rm s}}{N_{\rm m}\xi_{\rm I}}\frac{2\pi T_0}{I},
  \qquad  \mathfrak{T}_{\rm m}, \qquad  \frac{\xi_{\rm I}}{l_{\rm m}}, \qquad \frac{\xi_{\rm I}}{d_{\rm
  m}}.
\label{param}
\end{equation}
$N_{\rm s,m}$ are the densities of states at the Fermi energy in the
S, F. $\xi_{\rm s}$ is the length scale of Cooper pairs in S in
proximity systems (given by $\xi_{\rm s}=\sqrt{D_{\rm s}/2\pi T_{\rm
cS}}$ \cite{Radovic1}, with $D_{\rm s}$ the diffusion coefficient in
the S and $T_{\rm cS}$ critical temperature of the bulk S). The
magnetic stiffness length $\xi_{\rm I}=\hbar v_{\rm m}/2I$, where
$v_{\rm m}$ is the Fermi velocity and $I$ the exchange splitting in
F. $T_{0}$ is the BCS critical temperature of the SC and
$\mathfrak{T}_{m}$ is the parameter accounting for the interface
transparency \cite{Tag2}. $l_{m}$ and $d_{m}$ are the mean free path
and the thickness of the F layer respectively. Our strategy was to
fix as many as possible of these parameters with experimental
values.

\begin{figure}
  \includegraphics[width=8.0cm]{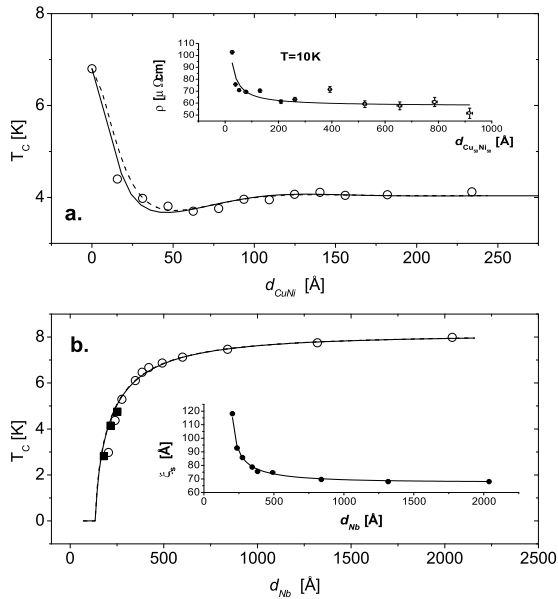}
\caption{(a) $T_{\rm C}$ vs. CuNi layer thickness, $d_{\rm
CuNi}$(=$d_{\rm m}$) for the S4 set. The lines are interpolated
curves assuming $\pi T \ll I$ (solid) and in the general case
(dashed), with parameters values of: $\xi_{\rm I}/l_{\rm m}$=2.8,
$\xi_{\rm s}/d_{\rm s}$=3.3, $\mathfrak{T}_{\rm m}$= 0.4, 0.45 and
$\varepsilon$=1.9, 2.3, respectively. In the inset the low
temperature resistivity of Cu$_{50}$ Ni$_{50}$ single layers is
plotted vs. $d_{\rm CuNi}$ (S2 set of samples). (b) $T_{\rm C}$
vs. Nb layer thickness, $d_{\rm Nb}$(=$d_{\rm s}$), in zero field
cooling for the samples of the S3 (circles) and S1 (squares) sets,
respectively. The lines are the same curves as in (a), plotted as
a function of $d_{\rm Nb}$, with $\xi_{\rm I}/d_{\rm m}$=1.18.
Inset: $\xi_{\rm s}$ vs. $d_{\rm s}$, the solid line is a fit of
the data with the assumption of the scaling: $\xi_{\rm s} \sim
T_{\rm C}^{-0.5}$.} \label{F8}
\end{figure}

The resistivity as function of thickness $d_{\rm m}$ for the CuNi
monolayers of S2 is plotted in the inset of Fig. \ref{F8}(a), along
with a fit to the Fuchs-Sondheimer relationship $\rho=\rho_{\rm
B}(1+3 l_{\rm m}/8d_{\rm m})$ where $\rho_{\rm B}$ is the bulk
resistivity \cite{Sond}. The fit yields $\rho_{\rm B}=$ 57 $\pm$ 1
$\mu\Omega$cm and $l_{\rm m}$= 44 $\pm$ 2 \AA. Resistance vs. $T$
measured during cooling in zero field was used to determine $T_{\rm
C}$ for the S3 set. The results are plotted in Fig. \ref{F8}(b). A
bulk $T_{\rm C}$ of $\approx 8$ K for Nb can be observed, lower than
the expected critical temperature of 9.2 K for pure Nb. The
discrepancy is due to residual impurities and other defects found in
thin film materials. In the limit of thin Nb layers, the proximity
effect becomes larger, resulting in a falling $T_{\rm C}$ until a
critical thickness $d^{\rm cr}_{\rm s} \approx 160$ \AA, below which
superconductivity is totally suppressed. The previous result is
consistent with reported data of Rusanov et al. \cite{Rusanov}. The
critical temperatures of the S1 samples are also plotted for
comparison. The points fall on the same curve as for S3, thus
showing that the layers outside the CuNi have no measurable effect
on the superconductor. On the same S3 set, the coherence lengths in
the out-of-plane field configuration were estimated. For this
purpose, the critical fields, $H_{\rm C \perp }$ vs. $T$ were
measured by sweeping $H$ at fixed $T$ and taking the value
corresponding to half-height of the resistance transition, averaged
over positive and negative fields. As predicted from Ginzburg-Landau
(GL) theory, $H_{\rm C \perp }$ goes linearly with $T$ near $T_{\rm
C}$, because of the 3D behavior expected in perpendicular field
configuration even for a thin film. From a linear fit of the curves
near $T_{\rm C}$, it is possible to extract the values of the GL
perpendicular coherence length $\xi_{\rm GL}$ by means of the
relationship:
\begin{equation}
H_{\rm C \perp}=\frac{\phi_0}{2\pi(\xi_{\rm
GL})^2}\left(1-\frac{T}{T_{\rm C}}\right)^2\label{Hc}
\end{equation}
where $\phi_0$ is the flux quantum. $\xi_{\rm s}$ is related to
$\xi_{\rm GL}$ by simple proportionality \cite{Koorevaar,Lazar},
$\xi_{\rm s}$=($2/\pi)\xi_{\rm GL}$.

From the inset of Fig. \ref{F8}(b), the value of $\xi_{\rm s}$ can
be obtained in the limit of thick Nb as $\xi_{\rm s}$=70 \AA. The
diagram also shows a divergent behavior of $\xi_{\rm s}$ in the
opposite limit of thin Nb. This can be understood within the
framework of the de Gennes-Werthamer theory of the proximity effect
\cite{Hauser}, where the coherence length of Cooper pairs in S is
calculated as proportional to $(\gamma_{\rm e} T_{\rm
C}/\sigma)^{-0.5}$, where $\sigma$ is the electrical conductivity
and $\gamma_{\rm e}$ is the coefficient of the electronic specific
heat. The solid line in the inset of Fig. \ref{F8}(b) is a fit
corresponding to the behavior $\xi_{\rm s}\propto T_{\rm C}(d_{\rm
s})^{-0.5}$. By using $\rho_{Nb}\approx$ 15 $\mu\Omega$cm, a value
of 5.9 mJ mol$^{-1}$ K$^{-2}$ for $\gamma_{\rm e}$ was calculated
from the fit, which is acceptably close to the known value for Nb of
7.8 mJ mol$^{-1}$ K$^{-2}$ \cite{Kittel}, given the approximations
we have made.

The $T_{\rm C}$ vs. $d_{\rm m}$  dependence for the S4 samples is
plotted in Fig. \ref{F8}(a). The samples are now capped with Ge
instead of Ta to avoid proximity with nonmagnetic metals for thin
F layers. A clear dip appears at $d_{\rm m}\approx$ 60 \AA. This
is in agreement with the data on Nb/CuNi bilayers reported in Ref.
\onlinecite{Fominov2002} (50 \AA) and Ref. \onlinecite{Rusanov}
(40 \AA), where higher Ni concentrations have been used. The
position of the minimum and $\xi_{\rm I}$ are related through:
$\xi_{\rm I}\approx 2 d_{\rm m}$ for $\mathfrak{T}_{m}>5$
\cite{Lazar} and can be acceptably extended to lower
$\mathfrak{T}_{\rm m}$ values, yielding in our case $\xi_{\rm I}
\approx 120$ \AA. The evaluation of $l_{\rm m}$, $d_{\rm m}$ and
$\xi_{\rm I}$ allows us to fix the values of the last two
parameters in (\ref{param}): $\xi_{\rm I }/l_{m} \simeq 2.8$ and
$\xi_{\rm I}/d_{\rm m} \simeq$ 1.18.

The remaining two parameters can be constrained by recalling that
for F/S trilayers $d^{\rm cr}_{\rm s}$ is related to
$\mathfrak{T}_{\rm m}$ by \cite{Lazar}:
\begin{equation}
\frac{d^{\rm cr}_{\rm s}}{\xi_{\rm s}} = 2\sqrt{2\gamma} \arctan
\Bigl( \frac{\pi}{\sqrt{2\gamma}}\frac{N_{\rm m} v_{\rm m}}{N_{\rm
s}v_{\rm s}}\frac{\xi_0}{\xi_{\rm
s}}\frac{1}{1+(2/\mathfrak{T}_{\rm m})} \Bigr),    \label{dcr}
\end{equation}
with $\gamma \simeq$ 1.7811 and $\xi_{0}$ the BCS coherence length
in the S, given by $\xi_{0}=(\hbar v_{\rm s})/(1.76\pi T_{0})$. The
latter relationship is valid in the limit of $d_{\rm m}\gg \xi_{\rm
I}$, but it is still a very good approximation for
$\mathfrak{T}_{\rm m}< 1$ or, more generally, in the regime where
$T_{\rm C}$ vs. $d_{\rm m}$ is nearly constant. By extracting
$N_{\rm s}/N_{\rm m}$ from (\ref{dcr}) and substituting into
$\varepsilon$ as given in (\ref{param}), we finally obtain:
\begin{equation}
\varepsilon\simeq \frac{3.9
\pi}{\sqrt\gamma}\frac{1}{1+(2/\mathfrak{T}_{\rm m})}\cot\Bigl(
\frac{d^{\rm cr}_{\rm s}}{2\sqrt{2\gamma}\xi_{\rm s}}\Bigr).
\label{edcr}
\end{equation}
The set of free parameters (\ref{param}) is then reduced to only the
interface transparency $\mathfrak{T}_{\rm m}$. The latter can be
found by interpolating the data in Fig. \ref{F8}(a), with the model
of Ref. \onlinecite{Tag2}, and using for $\varepsilon$ the values
given by (\ref{edcr}) as initial guesses. We then estimated
$\mathfrak{T}_{\rm m} \simeq 0.4$ and $\varepsilon \simeq 2$. In
order to check the consistency of our results, we calculated the
ratio $N_{\rm s}/N_{\rm m}$ for bulk epitaxial Cu$_{50}$Ni$_{50}$ by
means of the ASW software \cite{kubler}. By substituting $N_{\rm
s}/N_{\rm m}\simeq$ 0.25 into (\ref{param}) and using the definition
of $\xi_{\rm I}$ and (\ref{dcr}) we estimated $v_{\rm s} \approx3
\times 10^{7}$ cm s$^{-1}$ and $2I=1.8$ meV. The value of $v_{\rm
s}$ is fairly close to $2.77 \times 10^{7}$ cm s$^{-1}$ quoted by
Vodopyanov et al. \cite{Vodopyanov}, while the small exchange band
splitting can be qualitatively justified with the low Curie
temperature of the CuNi we used. This raises the concern that the
low value of $I$ brings the system close to the weak F/S limit ($I
\lesssim \pi T$). In our case ($T \approx 4$K) we have $I \approx
\pi T \approx 1$ meV. The data of Fig. \ref{F8}(a) were therefore
interpolated within the same model extended to the low $I$ region
\cite{Tag2}, represented by the dashed line in Fig. \ref{F8}(a). It
is reassuring to note that the values of the set of parameters
\ref{param} came out to be approximately the same as within the
previous approximation ($\varepsilon=2.3$, $\mathfrak{T}_{\rm
m}=0.45$), but the interpolation returned a higher value for $I$ of
8 meV, incompatible with the weak F/S limit. This, coupled with the
fact that the fitted curves are so similar, means that the concern
is not a serious one.

In the theory of Fominov et al. \cite{Fominov2002} the parameter
$\gamma_{\rm b}=RA/ \rho_{f} \xi_ {f}$ (where $R$ is the
resistance of an interface of area $A$ and $\rho_{f}$ and $\xi_
{f}$ are the resistivity and coherence length of the ferromagnet,
respectively) plays an analogous role to $\mathfrak{T}_{\rm m}$ in
the Tagirov picture. By means of the Landauer formula
\cite{Landauer1957}, $RA$ may directly linked to
$\mathfrak{T}_{\rm m}$ yielding: $\gamma_{\rm b}=2l_{\rm
m}/3\mathfrak{T}_{\rm m}\xi_{f}$ (the same relationship may be
obtained directly by comparing Eqs. (8) and (14) in Ref.
\onlinecite{Fominov2002} with Eqs. (6) and (14) of Ref.
\onlinecite{Tag1}, assuming a real valued diffusion coefficient).
Our measured value of $\mathfrak{T}_{\rm m}=0.4$ gives
$\gamma_{\rm b} = 0.6$ when this formula is used. Fominov et al.
\cite{Fominov2002} obtained a value of 0.3 when fitting their
theory to unpublished Nb/CuNi bilayer data of Ryazanov. Both
measurements agree that the transparency of a CuNi/Nb interface is
rather low.

\begin{figure}
  \includegraphics[width=8.0cm]{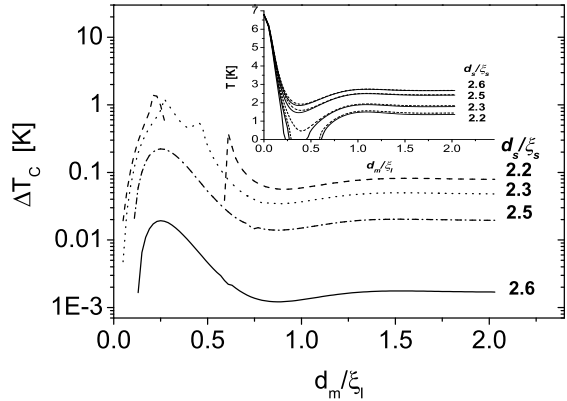}
\caption{ $T_{\rm C}$ shift vs. $d_{\rm m} / \xi_{\rm I}$. The log
scale is for better displaying the change of two orders of
magnitude in the $T_{\rm C}$ shift with changing $d_{\rm
s}/\xi_{\rm s}$. Inset: reduced $T_{\rm C}$  vs. $d_{\rm m}/
\xi_{\rm I}$ for the P (solid) and AP (dashed) cases used to
calculate the shift in the main graph. The parameters values are
$\xi_{\rm I}/l_{m}$=2.8, $\mathfrak{T}_{m}$=0.4 and
$\varepsilon$=2, which correspond to the measured values for
Nb/CuNi systems.} \label{F7}
\end{figure}

The $\Delta{T_{\rm C}}$ for Nb/CuNi systems can now be predicted as
a function of the thickness of S and F layers. The results are
summarized in Fig. \ref{F7}. As expected, $\Delta{T_{\rm C}}$
increases for smaller $d_{\rm s}/\xi_{\rm s}$, and has a clear
maximum for $d_{\rm m}/\xi_{\rm I} \sim$ 0.25. Actually, the real
shift is strongly affected by proximity with the layer external to
CuNi (mainly Py) for $d_{\rm m} \leq \xi_{\rm I}$, since $\xi_{\rm
I}$ can be regarded as the coherence length of the injected pair in
F \cite{Tag1}. In our case $\xi_{\rm I} \approx$120 \AA\, so only
the calculations with $d_{\rm m}\gg\xi_{\rm I}$ should be taken into
account. This can probably explain why $\Delta{T_{\rm C}}$ $\approx$
6 mK in Ref. \onlinecite{Gu} is two orders of magnitude smaller than
expected, for $d_{m}$=50 \AA. The result is that in the best case of
$d_{\rm s} \approx d^{\rm cr}_{\rm s}$, the largest shift possible
is $\Delta{T_{\rm C}}$ = 0.4 K. The maximum is accessible at
temperatures lower than 1 K and rapidly falls to 0.05 K after
0.1$d_{\rm s}/\xi_{\rm s}$, i.e. $d_{\rm s}$=7 \AA\ $+ d^{\rm
cr}_{\rm s}$. The region where the performance is enhanced is
therefore hardly accessible.

We have shown, by a series of experimental measurements of the key
physical parameters in the theory, that there are two main
limitations of the CuNi/Nb system as the core of a superconducting
spin-valve. These are the low value of interfacial transparency
$\mathfrak{T}_{\rm m} \approx 0.4$ (the same as for Pb/Fe systems
\cite{Lazar}, but far lower than in Nb/Gd systems, where
$\mathfrak{T}_{\rm m} \approx 1.75$ \cite{Vodopyanov}), and the high
ratio $d^{\rm cr}_{\rm s}/\xi_{\rm s}$ which is roughly 2. It is
desirable to have a superconducting layer that is much thinner than
its coherence length, but this is not possible because of the extra
pair breaking caused by the proximity effect in this system. By
calculating the $T_{\rm C}$ shift with the thickness corresponding
to the thinnest sample of S1, and considering an uncertainty of $\pm
5$ \AA\ in Nb thickness, we have $2.5 \leq d_{\rm s}/\xi_{\rm s}
\leq 2.6$, which yields $\Delta T_{\rm C}$ between 1 and 20 mK, in
accordance to the shift of 2.5 mK we measured experimentally.

\begin{acknowledgments}
We would like to thank K. Critchley and S. D. Evans for assistance
with the XPS measurements. This work was funded by the EPSRC, and by
the EU via project NMP2-CT-2003-505587 ``SFINx''.
\end{acknowledgments}


\bibliography{NbCuNiSpinSwitch}

\end{document}